\documentclass[longabstract,preprint,dvips,12pt]{aastex}

\usepackage[active]{srcltx}
\usepackage{natbib}
\usepackage{graphicx}
\usepackage[latin1]{inputenc}
\usepackage{amssymb}
\usepackage{amsmath}
\usepackage{xspace}
\usepackage{color}
\newcommand{\Ed}{E_{\rm d}}

\newcommand{\kt}{k_{\rm B}T}
\newcommand{\ktpar}{\kt_{\parallel}}
\newcommand{\ktperp}{\kt_{\hspace*{-0.15em}{\perp}}}
\newcommand{\HClp}{\ensuremath{{\rm HCl}^{+}}\xspace}
\newcommand{\HClpi}{\ensuremath{^1{\rm H}^{35}{\rm Cl}^{+}}\xspace}

\newcommand{\e}[1]{\ensuremath{\times 10^{#1}}}			

\title{Dissociative recombination measurements of HCl$^+$ using an ion storage ring
}

\author{O. Novotn\'y\altaffilmark{1}, 
	A.~Becker\altaffilmark{2},
	H.~Buhr\altaffilmark{2},
	C.~Domesle\altaffilmark{2},
	W.~Geppert\altaffilmark{3},
	M.~Grieser\altaffilmark{2},
	C.~Krantz\altaffilmark{2},
	H.~Kreckel\altaffilmark{2},  
	R.~Repnow\altaffilmark{2},
	D.~Schwalm\altaffilmark{2,4},
	K.~Spruck\altaffilmark{5},
	J.~St\"{u}tzel\altaffilmark{1},
	B.~Yang\altaffilmark{6,2}, 
	A.~Wolf\altaffilmark{2}, and
	D.~W.~Savin\altaffilmark{1}
	}

\date{\today }
\email{oldrich.novotny@mpi-hd.mpg.de}

\altaffiltext{1}{Columbia Astrophysics Laboratory, Columbia University, New York, NY 10027, USA}
\altaffiltext{2}{Max Planck Institute for Nuclear Physics, 69117 Heidelberg,
Germany}
\altaffiltext{3}{Department of Physics, Stockholm University, AlbaNova, SE-106 91, Stockholm, Sweden}
\altaffiltext{4}{Department of Particle Physics and Astrophysics, Weizmann
Institute of Science, Rehovot 76100, Israel}
\altaffiltext{5}{Institut f\"ur Atom- und Molek\"ulphysik, Justus-Liebig-Universit\"at Giessen, D-35392 Giessen, Germany}
\altaffiltext{6}{Institute of Modern Physics, Chinese Academy of Sciences, Lanzhou 730000, China}

\keywords{Astrochemistry --- Molecular data --- Molecular Processes --- Methods:
laboratory --- ISM: clouds --- ISM: molecules}

\begin{abstract}
We have measured dissociative recombination of \HClp with electrons using a
merged beams configuration at the heavy-ion storage ring TSR located at the Max
Planck Institute for Nuclear Physics in Heidelberg, Germany. We present the
measured absolute merged beams recombination rate coefficient for collision
energies from 0 to 4.5 eV. We have also developed a new method for deriving the
cross section from the measurements.  Our approach does not suffer from
approximations made by previously used methods.  The cross section was
transformed to a plasma rate coefficient for the electron temperature range from
$T=10$ to 5000~K. We show that the previously used \HClp DR data underestimate
the plasma rate coefficient by a factor of 1.5 at $T=10$~K and overestimate it
by a factor of 3.0 at $T=300$~K. We also find that the new data may
partly explain existing discrepancies between observed abundances of
chlorine-bearing molecules and their astrochemical models.
\end{abstract}

\begin{document}
\doublespace
\maketitle

\section{Introduction}
\label{introduction}
Despite the fact that the elemental abundance of chlorine is orders of magnitude
lower than that of more abundant elements such as carbon or oxygen,
astrochemists have been interested in chlorine chemistry for nearly 40 years
now. The evolving understanding of Cl in the interstellar medium (ISM) can be
followed in a series of papers over this time (e.g.,
\citealt{Jura:ApJ:1974,Dalgarno:ApJ:1974,Blake:ApJ:1986, Federman:ApJ:1995,
Amin:EMP:1996,Sonnentrucker:ApJ:2006,Neufeld:ApJ:2009, Lis:AA:2010,
de_luca:ApJL:2012, Neufeld:ApJ:2012}). A surprising development in this is that,
in spite of the low cosmic abundance of Cl, the latest observations indicate
that the abundances of chlorine-containing molecules can be comparable to those
of H$_2$O or CH \citep{Lis:AA:2010, Neufeld:ApJ:2012}. 
 
Until recently only one chlorine-bearing molecule had been observed in the ISM,
namely HCl (\citealt{Peng:ApJ:2010} and references therein). With the advent of
the {\it Herschel Space Observatory}, H$_2$Cl$^+$ and HCl$^+$ ions have now also
been detected \citep{Lis:AA:2010, de_luca:ApJL:2012}. This allows for a much
more detailed comparison of astrochemical models to the observations. The
H$_2$Cl$^+$ abundances derived from these observations are over ten times
greater than those of the latest models \citep{Neufeld:ApJ:2012}. Similarly, the
observed HCl$^+$ abundance is several times larger than predicted by models
\citep{Neufeld:ApJ:2009}. One possible candidate for explaining these
discrepancies would be an erroneous rate coefficient for
dissociative recombination (DR) of HCl$^+$ assumed in the models. For
a detailed description of the
Cl chemistry, see the review given by \cite{Neufeld:ApJ:2009}. To highlight the
importance of HCl$^+$ within the chlorine ISM chemistry, we discuss here a few
of the key reactions in the network. 

With a value of 12.97 eV the ionization potential of Cl lies
slightly below that of atomic hydrogen.
Thus, in the illuminated portions of interstellar clouds, photoionization of
neutral atomic H does not shield neutral atomic Cl from the interstellar
radiation field. As a result atomic chlorine is predominantly found in the form
of Cl$^+$. The uniqueness of chlorine astrochemistry is due to the exoergic
formation of HCl$^+$ via 
\begin{equation}
 {\rm Cl}^+ + {\rm H}_2 \rightarrow {\rm HCl}^+ + {\rm H}.
\end{equation}
All other elements predominantly found in interstellar clouds  as
singly charged atomic ions react endoergically with H$_2$. 
Once HCl$^+$ is formed it can react exoergically with H$_2$ to create
H$_2$Cl$^+$ via
\begin{equation}
 {\rm HCl}^+ + {\rm H}_2 \rightarrow {\rm H}_2{\rm Cl}^+ + {\rm H}.\label{eq_3}
\end{equation}
Neutral HCl is then formed, in part, by DR of H$_2$Cl$^+$ through
\begin{equation}
 {\rm e}^- + {\rm H_2Cl^+}  \rightarrow  {\rm H + HCl }.\label{eq_4}
\end{equation}
However, HCl$^+$ can also undergo DR via
\begin{equation}
 {\rm e}^- + {\rm HCl^+}  \rightarrow  {\rm H + Cl }.
\label{react:dr}
\end{equation} 
This reaction reduces the \HClp abundance, slowing  the formation of the
H$_2$Cl$^+$ and HCl via Reactions \ref{eq_3} and \ref{eq_4}, and thereby
affecting their equilibrium abundances.

To the best of our knowledge, there have been no previous investigations into DR
of HCl$^+$ until recently (e.g., our experimental work and exploratory
theoretical studies by \citealt{Larson:2012}). Lacking reliable data,
astrochemical models have used a ``typical'' diatomic DR rate coefficient for
that of HCl$^+$. It has been shown, however, that DR rate coefficients of
diatomics can differ by large factors from alleged typical values. For example,
for CF$^+$ \cite{Novotny:JpB:2005} measured a rate coefficient a factor of 4
lower than that commonly taken for the typical diatomic ion value. For this
reason \cite{Neufeld:ApJ:2012} have identified the DR rate coefficient of
HCl$^+$ as an ``urgently needed'' parameter significantly affecting the
uncertainty of chlorine-chemistry models.

To meet these needs, we have carried out DR measurements for HCl$^+$. The DR
pathways relevant for \HClp are described in Section~\ref{sec:DRpaths}. The
experimental setup, measurement method, and data analysis are discussed in
Section \ref{exp_setup} of this paper. In Section \ref{sec:res} we present the
resulting merged beams DR rate coefficient for \HClp, derive a DR cross section
and subsequently the  Maxwellian plasma DR rate coefficient. We discuss our
results and their implications for astrochemistry in Section \ref{sec:discuss}.
A summary is given in Section~\ref{sec:sum}.

\section{DR pathways for HCl$^+$} \label{sec:DRpaths}
In any DR process the incident electron couples within the Frank-Condon region
of the initial ionic state to form an excited neutral state
\citep{Larsson:book:2008}. This may autoionize back to form a molecular ion
again, or it can dissociate into neutral DR products. 

Depending on the nature of the excited neutral state into which the incident
electron is captured, two basic types of DR can be distinguished (e.g.,
\citealt{Larsson:book:2008}). In direct DR this neutral state is electronically
doubly excited and repulsive so that the molecule can directly dissociate. The
range of electron energies accessible for the transition within the Frank-Condon
region is given by the steepness of the neutral doubly excited state involved.
Usually this range is large and results in broad structures in the DR cross
section spectrum. Alternatively, the incident electron can be captured into a
bound neutral state. Such levels have discrete total energies and give rise to
sharp resonances in the energy dependence of the DR cross section as their
predissociation contributes to DR. This process is called indirect DR. The
dissociation in both DR types can induce multiple transitions between the
electronic states.

The neutral bound levels involved in indirect DR can often be grouped
into various Rydberg series. Members of a given series differ by the excitation
level $n$ of the captured electron. Each such series converges as $n \rightarrow
\infty$ to a single bound level of the ionic core with specific rotational,
vibrational, and electronic quantum numbers. In general, with
increasing collision energy for a given Rydberg series, the density of the
neutral states increases and then drops to zero above the series limit. DR
resonances from this indirect recombination process may be experimentally
unresolvable as they can overlap due to their natural widths and also due to the
limited experimental energy resolution. However, what may still be
experimentally distinguishable is a change in the DR cross section at the end of
a neutral Rydberg series, i.e., at the electron-ion collision energy
corresponding to the specific excitation of the ionic core. This change could be
either an increase or decrease, depending on the specific pathways involved as
well as potential quantum mechanical interferences \citep{Larsson:book:2008}.
The DR cross section may also drop above an ionic core excitation limit because
electron impact excitation of the ion competes with DR occurring at energies
above this limit.

Quantum interferences between neutral states belonging to various excitation
channels may also lead to unexpected structures in the DR spectrum.  More
discussion can be found, for example in \cite{Larsson:book:2008},
\cite{Wolf:JPCS:2011}, and \cite{Waffeu:PRA:2011}. Lacking detailed information
on these mechanisms for neutral levels lying in the \HClp electronic scattering
continuum, we here use the location of the ionic excitation thresholds and of
the main dissociation curves as a guide.

The basic properties of \HClp DR can be obtained from its thermochemistry. We
can calculate the exothermicity for DR of \HClp in the ground  ${\rm X}\,^2
\Pi_{3/2}$ state using the $\sim 13.6$~eV ionization energy of the hydrogen atom
and the $\sim 5.3$~eV proton affinity of chlorine \citep{NISTWebBookWhole}. For
collisions at $E=0$ this exothermicity results in an energy release of $\sim
8.3$~eV. This is less than the $\gtrsim 8.9$~eV and $\gtrsim 10.2$~eV  energies
needed to electronically excite Cl and H, respectively
\citep{AtomicSpectraDatabase}. Hence, low energy DR of \HClp results in both
atomic products being exclusively in their ground states. As the exothermicity
is insufficient to internally excite the DR products, it goes instead into the
kinetic energy released (KER), which is carried away by the products. The
channels with excited Cl, excited H, or both excited become accessible only for
$E\gtrsim0.6$~eV, $1.9$, and 2.5, respectively. 

DR of multielectron systems such as \HClp is a complex process that involves a
large number of resonances in the electron scattering continuum.  The molecular
structure of \HClp and HCl determine the behavior of the DR cross section versus
electron-ion collision energy $E$. The most likely dissociative route from
potential energies near the \HClp ground state toward the H and Cl ground states
is via the HCl repulsive $^3\Sigma^+$ potential curve. This curve crosses the
\HClp ground state potential curve at internuclear distances somewhat smaller
than the Franck-Condon region of the $\HClp(v=0)$ initial state
\citep{Bettendorff:CP:1982}. Thus, low-energy DR here is most likely driven by
the indirect process. In the first step the electron is captured to one of the
ro-vibrationally excited neutral levels converging to an excited $\HClp$ core.
This then couples to the $^3\Sigma^+$ repulsive potential curve and
predissociates. Data on molecular levels formed by optical excitation of HCl to
states near (but below) the ionization threshold for forming \HClp are available
from spectroscopic and theoretical studies (e.g.,
\citealt{Bettendorff:CP:1982,Liyanage:JCP:1995,Alexander:CP:1998,
Romanescu:JCP:2007,Lefebvre-brion:JCP:2011,Long:JCP:2013}).

Based on the above discussion, structures in the energy dependence of the DR
cross section may occur near \HClp excitation thresholds. For ground state ${\rm
X}\,^2\Pi_{3/2}$ \HClp, some of these limits are rotational excitations, fine
structure ${\rm X}\,^2\Pi_{3/2} \to\,{\rm X}\,^2\Pi_{1/2}$ excitation, and
vibrational excitations. Another series worth mentioning may arise from
electronic excitation of \HClp into the ${\rm A}^2\Sigma^+$ state.

Electron-induced processes that could compete with DR and thus reduce the DR
signal are all endothermic. One of these is ion pair formation which can yield
H$^+$+Cl$^-$ or H$^-$+Cl$^+$ at energies above 1.7~eV and 3.9~eV, respectively.
Another electron-driven process is dissociative excitation (DE) forming H +
Cl$^+$ or H$^+$ + Cl.  This reaction is endothermic by $\sim4.7$~eV and
$\sim5.3$~eV, respectively.

\section{Experimental}
\label{exp_setup}
\subsection{Setup}\label{sec:setup}
Measurements were carried out at the TSR heavy ion storage ring of the Max
Planck Institute for Nuclear Physics in Heidelberg, Germany \citep{Habs1989}.
With a base pressure of $\sim 10^{-11}$~mbar, TSR  is ideally suited to simulate
the two-body collision regime important for interstellar gas-phase chemistry.
Details on various aspects of the merged beams technique as used at the TSR and
the recent developments of the photocathode electron beam and the fragment
imaging technique have been described by \cite{Amitay:PRA:1996},
\cite{Krantz:JPCS:2009}, and \cite{Novotny:JPhysChemA:2010}. Here we discuss
only those aspects specific to this study.

\HClp ions were produced in a cold-cathode Penning ion source from a mixture of
H$_2$ and HCl parent gases. Ions were extracted from the source and brought to a
kinetic energy of 2.4~MeV using a Pelletron-type accelerator. The magnetically
mass-filtered beam of \HClpi was then selected for and injected into the TSR
where it was stored for $\sim33$~s.

The stored ion beam was merged for $\sim 1.5$~m with a nearly monoenergetic
electron beam. This beam, which we refer to as the Target, is generated by a
photocathode operated at temperatures of around 100 K \citep{Orlov:NIMA:2004,
Orlov:JPCS:2005, Orlov:JAP:2009}, thereby enabling us to perform electron-ion
collision studies with high energy resolution \citep{Sprenger:NIMPRA:2004}.
After ion injection at time $t=0$ and continuing until $t = 16$~s, the electron
beam was velocity matched to the ions. During this time elastic collisions of
the ions with the low energy spread (i.e., cold) electron beam transferred
energy from the circulating ions to the single-pass electrons. This mechanism of
electron cooling of the ions, also known as phase-space cooling
\citep{Poth:PR:1990}, results in reducing the ion beam velocity spread
and a reduced ion beam diameter.

The ion beam energy of 2.4~MeV was limited by the highest available magnetic
field strength available for deflecting the stored ions along the orbit within
TSR. Velocity matching the electrons with the 36 amu \HClp ions resulted in an
electron beam energy as low as 36.9~eV. Under these conditions, obtaining a
phase-space cooled ion beam requires an electron beam of both high  density and 
low temperature. This was accomplished  by utilizing the cryogenically cooled
photocathode electron source in the Target which provided an electron density of
$n_{\rm e}\approx 2\e{6}$~cm$^{-3}$  and a low energy spread as we discuss
below. The cooling capability of the Target for high-mass ions was demonstrated
earlier for a beam of 31 amu CF$^+$ ions (\citealt{Krantz:JPCS:2009}). In the
present measurement, in spite of the low electron energy and high ion mass, an
ion beam diameter as low as 0.5 mm was obtained. The second electron beam device
of the TSR, referred to as the Cooler \citep{Steck:NIMA:1990}, proved to have
insufficient cooling force to cool such heavy, slow molecular beams within the
available storage time. Therefore the Cooler was not used for the results
presented here.

After cooling, the Target electron beam was used as the interaction medium to
measure \HClp DR for a range of electron-ion center of mass collision energies
by varying the laboratory energy of the electron beam. Neutral DR products
generated in the Target were not deflected by the first dipole magnet downstream
of its position in the TSR and continued ballistically until they hit a
detector. Data were collected with this detector from $t = 16$~s to $t = 33$~s
as described in Section~\ref{sec:meas}. 

The energy spread of the photocathode-generated electron beam is parametrized by
effective temperatures $T_\perp$ and $T_{||}$, perpendicular and parallel to the
bulk electron beam velocity, respectively. The magnetic field guiding the
electrons was higher at the photocathode, as compared to the interaction zone,
by a factor of $\xi = 20$ for most of the measurements. This expansion of the
magnetic field leads to a lowering of the perpendicular electron temperature
\citep{Danared:NIMA:1993,Orlov:JPCS:2005}. By fitting  dielectronic
recombination resonances \citep{Lestinsky:prl:2008} and from fragment imaging
spectra obtained at similar values of $\xi$ \citep{Stuetzel:dis:2011}, we deduce
for the present measurement effective temperatures of  $\ktperp=1.65\pm0.35$~meV
and $\ktpar= 25^{+45}_{-5}~\mu$eV, where $k_{\rm B}$ is the Boltzmann constant.
Here and throughout all uncertainties are quoted at an estimated $1\sigma$
statistical confidence level. Some of the data were also acquired with an
expansion factor of $\xi = 40$, for which a reduced transverse electron
temperature of $\ktperp \approx 0.83$~meV is estimated, while $k_{\rm B}T_{||}$
is expected to remain unchanged (e.g., \citealt{Danared:NIMA:1993}). The
relevant temperatures $\ktperp$ and $\ktpar$ enter the data analysis as
described in Section~\ref{sec:PlasmaRateCalc}.

The interaction geometry is determined by the shapes of the overlapping electron
and ion beams. The electron beam geometry is determined by the magnetic fields
in the Target \citep{Sprenger:dis:2003}. The electron beam can be treated as a
cylindrical body, bent at the ends, with a diameter $d_{\rm e}$ given by the
product of the cathode size and the square root of the expansion factor $\xi$.
This yields $d_{\rm e} = 12.6$~mm and 17.8~mm for $\xi = 20$ and 40,
respectively. The cooled ion beam diameter $d_{\rm i}$ is typically less than
0.5~mm. The electron and ion beam geometries are shown in Figure \ref{fig:geom}
and are discussed in more detail in Appendix~\ref{app:A}.

The fact that the electron beam is significantly wider than the ion beam enabled
the complete spatial overlap of the ion beam by the electrons in the interaction
region. We optimized the alignment of the beams by minimizing the time needed
for phase-space cooling of the beam \citep{Hochadel:NIMA:1994}.
Better alignment gives a better overlap, thereby reducing the required cooling
time.  Observationally, we monitored the necessary cooling time and beam size
either using a beam profile monitor \citep{Hochadel:NIMA:1994} or by examining
the center-of-mass for DR events as projected on the MCP-imaging detector
\citep{Amitay:PRA:1996,Krantz:JPCS:2009}. 

The bulk electron beam energy in the laboratory frame, $E_{\rm meas}$, is needed
to calculate the center of mass electron-ion collision energies during the
measurement. We determine $E_{\rm meas}$ from the measured cathode voltage by
correcting it for the space charge of the electron beam \citep{Kilgus:PRA:1992}.
This correction requires knowing the electron density profile, which we measured
with the method described in \cite{Sprenger:NIMPRA:2004}.
The electron density is expected to be homogeneous along the beam axis.

\subsection{HCl$^+$ internal excitation}
The \HClp ions produced in the discharge are expected initially to possess
electronic, vibrational, rotational, and fine-structure excitation.  We estimate
that much of this internal excitation radiatively relaxes during the $16$~s
of electron cooling.  Electronically, the only known \HClp metastable state
below the dissociation limit is the A$\,^2\Sigma^+$ which lies $\sim3.6$~eV
above the X$\,^2\Pi$ ground state. Measured and calculated radiative lifetimes
for A$\,^2\Sigma^+\rightarrow {\rm X}\,^2\Pi$ transitions are shorter than
$\sim3.4~\mu$s (\citealt{Pradhan:JCP:1991} and references therein). 
The known radiative lifetimes for vibrational relaxation of the X$\,^2\Pi$
electronic ground state span from $\sim4.9$~ms for
$v=1\rightarrow0$ to $\sim1.3$~ms for $v=9\rightarrow8$
\citep{Pradhan:JCP:1991}. Higher vibrational levels are expected to
decay even faster. Thus we expect the stored ions to cascade quickly to their
ground electronic and vibrational levels.

We predict that the rotational lifetimes are sufficiently short so that during
the initial phase most of the rotations $J$ will have radiatively relaxed and
come into equilibrium with the $\sim300$~K black-body radiation of the vacuum
chamber. We have calculated the rotational radiative lifetimes of
X$\,^2\Pi_{3/2}(v=0)$ for levels ranging from $J=3/2$ to $J = 51/2$. In
actuality, we could truncate the calculations at $J=20/2$, as the higher levels
decay so rapidly that they do not affect the precision of the model. Our
approach uses a method similar to that of \cite{Amitay:PRA:1994}. The static
dipole moment for \HClp was taken from \cite{Cheng:PRA:2007}. The dominant
emission lines are expected to be $J\rightarrow J-1$. Restricting ourselves to
these transitions, we have built a relaxation model using the spontaneous
radiative decay lifetimes along with stimulated emission and absorption by the
300~K black-body radiation. For the initial rotational excitation we have taken
a Boltzmann distribution at a temperature of 8000~K. This is approximately the
excitation temperature derived for CF$^+$ produced in the same ion source
\citep{Novotny:JPCS:2009}. After the initial 16~s of ion storage, the predicted
average excitation energy exceeds the 300~K equilibrium by only 10\%. The
excitation energy averaged over the ion population during the $t=16-33$~s
measurement window  exceeds room temperature excitation by only 3\%. This
predicted level of excitation might be slightly overestimated due to the
omission of spin-orbit coupling which may provide extra decay pathways. We have
also not accounted for the possible additional acceleration of the rotational
cooling from super-elastic ion collisions with electrons (e.g.,
\citealt{Shafir:PRL:2009}). 

The one excitation which is unlikely to relax during ion storage is the fine
structure splitting of the X$\,^2\Pi_{3/2-1/2}$ which amounts to $\sim80$~meV
\citep{Sheasley:JMS:1973}. We are unaware of any published lifetime estimates
for this transition. However, for a rough estimate we can use the radiative
lifetime of the fine-structure excited $J=1/2$ level in the isoelectronic atomic
system Ar$^+$ with an energy of $\sim 165$~meV. That lifetime is calculated to
be 23.7~s \citep{NIST_MCHF}. Since the radiative decay rate of this magnetic
dipole transition scales as the third power of the transition frequency, then
leaving aside the details of the transition matrix element, we expect that the
lifetime of the corresponding level in the \HClp molecular state will be longer
by almost an order of magnitude. Hence, the X$\,^2\Pi_{1/2}$ radiative lifetime
is expected to significantly exceed the ion storage time here.

\subsection{Measurement procedure}\label{sec:meas}
Data acquisition began after the 16~s period of injection and cooling were
completed. During data acquisition the Target beam energy was stepped repeatedly
between {\it cooling}, {\it measurement}, {\it reference}, and {\it off} (all
defined below).  The time durations were 35, 25, 25, and 25 ms, respectively. An
additional 5~ms delay before each step was added to allow the power supplies to
reach the desired voltage. This cycle was repeated 130 times for a total of
$\sim 17$~s. 

For the first step, cooling, the electrons were
velocity matched to the ions at an electron energy of $E_{\rm cool}=36.91$~eV in
the laboratory frame. These interleaved cooling steps ensured a constant
phase-space spread of the ions during the data taking. 

In the second step, measurement, the electron beam energy was detuned,
giving a mean energy of $E_{\rm meas}$ in the laboratory frame.
The nominal center of mass collision energy can then readily be calculated from
the mean electron and ion velocities in the laboratory frame. This yields what
we call the detuning energy 
\begin{equation}
E_{\rm d} = \left(\sqrt{E_{\rm meas}} - \sqrt{E_{\rm cool}}\right)^2. 
\label{eq:ed}
\end{equation}
 The detector count rate in this step was used to determine the
merged beams rate coefficient versus $\Ed$. The detuning energy was changed
for each new ion injection. 

The third step, reference, was included as a cross check and for normalization.
The energy in this step was set to a constant value of $\Ed=0.019$~eV. The
resulting signal was used to monitor the ion beam intensity. 

In the last step, off, the electron beam was not admitted into the interaction
region. The resulting background signal was due solely to ion interactions with
the residual gas inside TSR.

DR events were measured using a $10\times10$~cm$^2$ Si surface-barrier detector
located $\sim12$~m downstream of the Target. Fragments from a given dissociation
event arrived at the detector with only a few nanosecond difference in flight
times. This is shorter than the detector time resolution and so only one pulse
was counted for each dissociation event, independent of the number of products
reaching the detector. The high exothermicity of \HClp DR results in relative
fragment kinetic energies of up to ${\rm KER} \approx 8.3$~eV. This can cause a
large displacement between the fragments when reaching the detector. The
resulting positions of some of the H~fragments exceeded the size of the
detector. The heavier Cl~fragments, however, were confined to a narrow cone and
100\% of the DR-generated Cl struck the detector. We verified this using an
MCP-imaging technique (e.g., \citealt{Amitay:PRA:1996}). Thus, to achieve
essentially 100\% DR event counting efficiency, we derive our signal from the
count of detector events independent of the number of fragments detected in each
of them.

The detector count rate consisted of both DR and background events. The
dominant source of this background were collisions of the ions with the
residual gas. We have corrected for these events by taking the count rate at
measurement and subtracting the count rate acquired with the electron beam off.
Our approach does not account for potential background due to non-dissociative
recombination, but this is expected to be negligible at the energies studied
\citep{Krauss:ApJL:1973, Dalgarno:RPP:1976, Larsson:book:2008}. 
Additional background due to electron-driven DE
forming ${\rm H^+ + Cl}$ was not an issue as the measured collision energies
were all significantly below the DE threshold of 5.3~eV.

The measured relative merged beams DR rate coefficient is determined by
normalizing the recorded DR signal by the electron density and ion number (e.g.,
\citealt{Amitay:PRA:1996}). The electron beam density is well detemined by the
measured beam current, energy, and geometry \citep{Sprenger:NIMPRA:2004}.
However, the maximum stored \HClp ion current of $I_{\rm i}<1~\mu$A was too low
to be determined directly using a DC current transformer. Instead we used a
measurement of the relative ion beam intensity by taking the detector signals
from the reference and off steps as proxies of the relative ion beam intensity.
This allows us to normalize our measured data to obtain a relative merged beams
DR rate coefficient versus $\Ed$. The whole curve was then scaled using an
absolute measurement of the merged beams DR rate coefficient at matched electron
and ion velocities (i.e., $\Ed=0$~eV). Here we used an independent method based
on precise ion beam storage life time measurements with and without the electron
beam present in the interaction zone \citep{Novotny:APJ:2012}. This method
requires using the length of the beam interaction region, which was determined
from the known beam geometries to be $L=1.570$~m. 

In the past, researchers have corrected the merged beams DR rate coefficient for
the effects from the merging and demerging of the electron beam with the stored
ions \citep{Lampert:PRA:1996}. Below we introduce a new data analysis method
which avoids the necessity of this correction.

\subsection{Generating a Plasma DR Recombination Rate Coefficient}
\label{sec:PlasmaRateCalc}
Using storage ring results one can generate a DR rate coefficient suitable for a
plasma with a Maxwell-Boltzmann collision energy distribution. This involves
deconvolving the measured DR data to remove the effects of the experimental
electron velocity spread and the beam overlap geometry. \cite{Mowat:PRL:1995}
and \cite{Lampert:PRA:1996} have presented approximate methods for addressing
these two issues. The resulting cross section can then be convolved with a
Maxwell-Boltzmann distribution to generate a plasma rate coefficient suitable
for astrochemical modelling. Here we have developed a more precise method for
deriving the cross section which avoids making the approximations used by
\citeauthor{Mowat:PRL:1995} and \citeauthor{Lampert:PRA:1996}

\subsubsection{Measured Merged Beams DR Rate Coefficient}
\label{sec:MBrate}

Experimentally we have measured a merged beams DR rate coefficient $\alpha_{\rm
mb}$ which is given by
  \begin{equation}
  \alpha_{\rm mb}(\Ed) = \int \sigma(E)\,v\,f_{\rm
  mb}(E, \Ed, T_{||}, T_\perp, \boldsymbol{X})\,dE.\label{eq:ratecross}
  \end{equation}
Here $\sigma$ is the DR cross section; $E$ and $v$ are the center of mass
electron-ion energy and the relative velocity, respectively; and $f_{\rm mb}$ is
the center of mass energy spread taking into
account both the electron beam energy spread and the experimental geometry. 
The relationship between the collision energy and velocity is given by
$E=\frac{1}{2}\,\mu\,v^2$, where $\mu$ is the reduced mass of the collision
system. The large mass difference between the electron and ion allows us to
set $\mu = m_{\rm e}$. The term $f_{\rm mb}$ is a function of the detuning
energy; the electron energy spread, which is given by the effective
temperatures $T_\perp$ and $T_\|$; and the overlap geometry between the beams,
symbolically represented as~$\boldsymbol{X}$.

For parallel ion and electron beams the velocity spread can be described by a
``flattened'' Maxwellian distribution in the center of mass frame. This velocity
distribution is discussed in more detail by \cite{Andersen:prl:1989} and
\cite{Poth:PR:1990}, while the corresponding energy distribution function is
described by \cite{Schippers:AA:2004}. Note, that a pre-factor of $\frac{1}{2}$
is missing from equation (1) given in Schippers et al.

In the merging and de-merging regions of the Target, the energy distribution is
additionally distorted due to the beams no longer running co-linearly.
\cite{Lampert:PRA:1996} discusses the corresponding increase of collision
energies, assuming that $T_\perp=T_{||}=0$. We are not aware, however, of an
analytical representation of $f_{\rm mb}$ which accounts for the effects of both
the velocity spread and the full overlap geometry. To address this issue, we
have developed a numerical method for describing such an energy distribution
which we discuss below.

\subsubsection{Limitations of Previous Methods for Extracting a DR Cross
Section}
\label{sec:crossecder}
The traditional methods for deconvolving the cross section from the merged beams
rate coefficient first correct the measured data for the overlap geometry
effects using, for example, the method of \cite{Lampert:PRA:1996}. 
The corrected data are then treated as if the beams are parallel with an energy
distribution
$f_{\rm mb}^*$, giving
\begin{equation}
\alpha_{\rm mb}^*(\Ed) = \int \sigma(E)\,v\,f_{\rm
mb}^*(E, \Ed, T_{||}, T_\perp)\,dE.\label{eq:ratecrosscorr}
\end{equation}
The analytic form of $f^*_{\rm mb}$ is given by \cite{Schippers:AA:2004} and
the characteristic experimental energy spread in $f_{\rm mb}^*$ 
corresponds to \citep{Muller:PT:1999} 
\begin{equation}
\Delta E \approx \sqrt{(\ktperp\,\ln{2})^2+16
\ln{2}\;\ktpar\,\Ed}.\label{eq:res}
\end{equation} 
For high energies, where $\Ed \gg
\ktperp$, the relative width $\Delta E/\Ed$ decreases to a small value. In
that case, $f_{\rm mb}^*$ can often be approximated by a delta function
$\delta(E-\Ed)$.
Equation (\ref{eq:ratecrosscorr}) then collapses to $\alpha_{\rm mb}^*(\Ed)
=\sigma(\Ed)\,v$, and the cross section simplifies to
\begin{equation}
\sigma(\Ed) = \alpha_{\rm mb}^*(\Ed) \sqrt{\frac{m_{\rm e}}{2\Ed}}.
\label{eq:sigmasimple}
\end{equation}

The situation is not so simple for $\Ed\lesssim\ktperp$. This regime is
particularly critical for obtaining an accurate plasma rate coefficient at the
low temperatures relevant to the cold ISM. At these low detuning energies the
cross section can be approximately derived using a method introduced by
\cite{Mowat:PRL:1995}. Their technique makes use of the fact that typically
$\ktperp\gtrsim50\,\ktpar$ in ion storage ring experiments. They go on to
neglect the longitudinal energy spread by setting $\ktpar = 0$. This makes
$\Delta E$ and the shape of $f_{\rm mb}^*$ independent of $\Ed$ and one can then
deconvolve $\alpha_{\rm mb}^*$ using Fourier analysis.

\begin{figure}
    \plotone{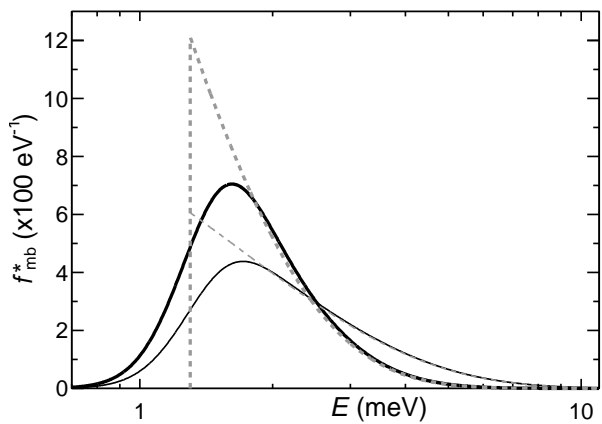}
    \caption{\doublespace
The collision energy distribution $f_{\rm mb}^*$ for parallel
electron and ion beams at a detuning of energy $\Ed=1.3$~meV. The black full
curves use the Target longitudinal electron beam energy spread of $\ktpar=25~
\mu$eV. The gray dashed curves are for $\ktpar=0$. The thin curves assume a
transverse electron energy spread of $\ktperp=1.650$~meV, while the thick curves
use $\ktperp=0.825$~meV. 
	\label{fig:fexp}
	}
\end{figure}

There are, however, a number of issues which the \citeauthor{Mowat:PRL:1995}\
approach raises. Their method ignores the fraction of electron-ion collision
energies $E$ lying below $\Ed$. This can be seen in Figure \ref{fig:fexp} which
shows the energy distribution function $f_{\rm mb}^*$ using two sets of Target
electron temperatures. As a representative detuning energy we have chosen $\Ed =
1.3$~meV, which corresponds to a typical particle energy at molecular cloud
temperatures of $10$~K. In Figure \ref{fig:fexp}, we also plot $f_{\rm mb}^*$
for the two Target values of $\ktperp$ but with $\ktpar=0$. These energy
distributions clearly differ from those using a realistic $\ktpar=25~\mu$eV. In
$f_{\rm mb}^*(\ktpar = 0)$ there is no collision energy population below $\Ed$.
For $f_{\rm mb}^*(\ktpar = 25 \mu$eV$)$ and an assumed value of
$\ktperp=1.65$~meV (0.825~meV), about 5\%~(10\%) of the distribution is shifted
to below $\Ed$. This fraction increases for decreasing $\ktperp/\ktpar$ or with
increasing $\Ed$, further reducing the validity of setting $\ktpar=0$ in order
to extract cross sections. Additionally, the potential errors in this method may
increase if the cross section at low energies is highly structured or increases
rapidly with decreasing energy. By using an energy distribution function which
does not accurately represent the experimental conditions, the deconvolutions
can accordingly under- or over-estimate the cross section at a given value of
$\Ed$.

\subsubsection{New Method for Extracting a DR Cross Section}
\label{sec:crossecdernew}
The approach we use here is to create an empirical model cross section spectrum
$\sigma'(E)$ and, following equation (\ref{eq:ratecross}), convolve it with
$f_{\rm mb}$ to generate a model experimental rate coefficient $\alpha'_{\rm
mb}$. We represent $\sigma'(E)$ using a histogram-shaped function with an energy
binning comparable to the energy resolution $\Delta E$ at $E$  as given by
equation (\ref{eq:res}) with $E=\Ed$. The amplitudes within each bin are treated
as free fitting parameters and adjusted iteratively to minimize the $\chi^2$
between the model $\alpha_{\rm mb}'$ and the measured $\alpha_{\rm mb}$. The
combined effects of the electron beam energy spread and the overlap geometry are
treated using a Monte Carlo simulation of $f_{\rm mb}$ for the integration of
equation (\ref{eq:ratecross}). In some cases (especially when the bin size
becomes smaller than the energy resolution) the resulting $\sigma'(E)$ can
fluctuate significantly from one to the next bin due to numerical instabilities.
The corresponding uncertainties, however, essentially cancel out when
integrating the cross section to generate a plasma rate coefficient as described
below. This method accounts for both the electron energy spread and the beam
overlap geometry. Thus our deconvolution method does not require the measured
merged beams rate coefficient be corrected for the overlap geometry, nor is it
based on the assumption that $\ktpar=0$.  In contrast to the traditional methods
we thereby largely avoid the uncertainties introduced by those assumptions.
Technical details of our method are discussed at length in
Appendix~\ref{app:A}. 

\subsubsection{Derived Plasma DR Rate Coefficient}
\label{sec:plasmader}
Using the extracted $\sigma'(E)$ as a representation of $\sigma(E)$, we can 
generate a rate coefficient for a plasma with a thermal Maxwell-Boltzmann
distribution $f_{\rm pl}$. The resulting rate coefficient as a function of the
plasma temperature $T$ is given by
\begin{equation}
\alpha_{\rm pl}(T) = \int \sigma(E)\,v\,f_{\rm pl}(E, T)\,dE.\label{eq:rateplas}
\end{equation}
The analytical form of $f_{\rm pl}(E, T)$ is well known and numerical
integration of equation (\ref{eq:rateplas}) is straightforward.

\subsection{Uncertainties in the measured merged beams DR rate coefficient}
The dominant uncertainty in the measured merged beams DR rate coefficient
derives from the ion beam storage life time measurements used to put the
experimental results on an absolute scale.  This error amounts to 11\%.  
Additional minor sources of error arise from the
electron density evaluation ($\sim 4\%$) and the uncertainty on the electron
beam geometry within the Target ($\lesssim 2\%$). The total systematic
uncertainty for the absolute scaling is then 12\%.

The statistical uncertainty of each data point is given by the counting
statistics for the number of signal and background counts.
At collision energies $\Ed<10$~meV this error amounts to $\sim 5$\%. At higher
energies this error increases as the signal and background become comparable.
The statistical error is $\sim45$\% at $\Ed\approx0.1$~eV and grows up to
$\sim200$\% at $\Ed\approx1$~ eV.

\section{Results}
\label{sec:res}
\subsection{Merged Beams Recombination Rate Coefficient}
\label{sec:resrate}

Figure \ref{fig:rateexpA} presents our merged beams rate coefficient for DR of
\HClp. The error bars show the 1$\sigma$ counting
statistics. The detuning energy ranges from $\Ed=12~\mu$eV to 4.5~eV. The
data were acquired with $\xi=20$ and cover the storage times of $16-33$~s.

Although our model for the rotational relaxation of HCl$^+$ discussed in Section
\ref{sec:setup} indicates that most of the ions are in the thermal equilibrium
with the black body radiation of the TSR chamber, a small amount of rotational
relaxation may still occur during the measurement period. To test for any effect
of this cooling we split the acquired data into two sets covering the storage
times of $16-24$~s and $24-33$~s and analyzed these data sets independently. DR
rate coefficients typically depend on the rotational excitation of the parent
ions \citep{Larsson:book:2008}.  Hence any significant change in rotational
population should be observable in the DR rate coefficient spectra. However, to
within their statistical uncertainties, the resulting two data sets for
$\alpha_{\rm mb}$ were equal at all values of $\Ed$. This strongly suggests that
the rotational excitation of the ions reached equilibrium with the black body
radiation of the vacuum chamber during the initial electron cooling phase.

A possible source of concern in storage ring merged beams experiments is that
the low energy $\alpha_{\rm mb}$ can potentially be distorted if the ion beam
energy is dragged by the detuned electron beam. The derivation of $\Ed$ through
equation (\ref{eq:ed}) rests on the assumption of a constant mean ion beam
velocity which is matched to the mean electron beam velocity at cooling. This
assumption breaks down if during the measurement step the ion beam is dragged
towards the detuned electron beam velocity by dynamical friction forces of the
same type as those causing electron cooling \citep{Poth:PR:1990}. The actual
collision energy is then smaller than that deduced from equation~(\ref{eq:ed}).
Since low energy DR for HCl$^+$ increases with decreasing energy this will
enhance the measured signal at higher energies over what one would expect if
there were no dragging. This will appear in the data as a broadening of
$\alpha_{\rm mb}$ close to 0~eV. In order to exclude such an experimental
artifact we have performed a series of tests described below. Based on these
studies, we believe the ion beam dragging is not an issue in our experiment and
that the low energy shape in $\alpha_{\rm mb}$ is due solely to DR.

The stability of the ion beam energy was tested by varying the duration of the
various steps in the data cycle. The standard data cycle was chosen to optimize
the duty factor and cooling as well as reduce dragging effects. All of the tests
in this paragraph further improved cooling and reduced any potential dragging
effects, but at the expense of significantly decreasing the measurement duty
factor. The cooling step was lengthened by a factor of $\sim3$ to provide
additional cooling and improve the ion beam stability. The measurement step was
shortened from 25~ms to 10~ms to reduce any dragging effects. Additionally we
varied the energy difference between the reference step and the cooling
step. If cooling were insufficient this would help to further minimize the
dragging of the ion beam energy when at reference. To within the statistical
accuracy of the measurement, all of these tests produced an $\alpha_{\rm mb}$
that agreed with the data obtained under standard measurement conditions. Thus
dragging of the ion beam energy does not appear to contribute to the observed
low energy rate coefficient.

\begin{figure}
    \plotone{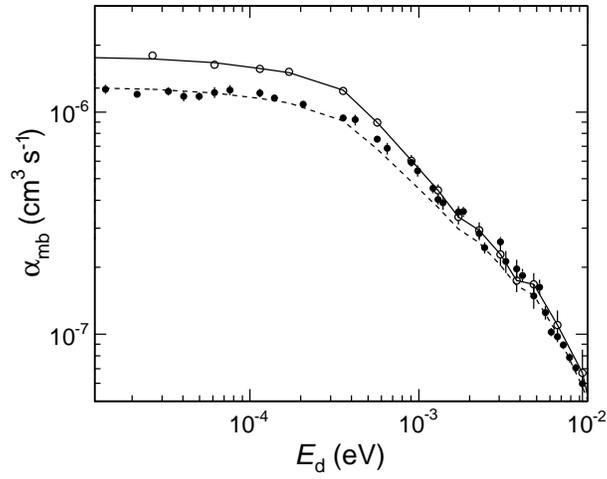}
    \caption{\doublespace
The experimental HCl$^+$ DR rate coefficients for
adiabatic magnetic expansion factors of $\xi = 40$ and $\xi = 20$ are plotted
by the open and full circles, respectively. The solid line shows the
fit $\alpha'_{\rm mb}(\xi=40)$ while the dashed line is the converted
$\alpha^{''}_{\rm mb}(\xi=20)$ based on $\xi=40$ data. See Section
\ref{sec:resrate} for details.
	\label{fig:fexp40}
	}
\end{figure}

To further test the DR origin of the low energy data, we have measured the
behavior of $\alpha_{\rm mb}(\Ed)$ for two different energy spreads. As one
reduces the width of the collisional energy distribution this results, after
averaging over the electron energy distribution, in a higher amplitude for
$\alpha_{\rm mb}$ at $E_{\rm d}\sim0$~eV. Dragging, on the other hand, modifies
only the width of the low energy DR spectrum but not the peak amplitude. Here we
narrowed the experimental energy spread by increasing the adiabatic expansion
factor of the magnetic guiding field in the Target electron gun  from $\xi=20$
to $\xi=40$. The tranverse temperature is expected to scale as $\xi^{-1}$. 
Figure \ref{fig:fexp40} presents both the $\xi =20$ and $\xi = 40$ results.  The
higher expansion data do indeed show an additionally increased amplitude of the
rate coefficient at $\Ed \lesssim 1$~meV. The half width half maximum of the
data below $\sim 1$~meV also narrowed by $\sim17$\%. This strongly
suggests that the observed low energy rate coefficient is caused by the DR 
process and not by beam dragging artifacts.

Still, the higher amplitudes in the low energy $\alpha_{\rm mb}(\xi = 40)$ data
do not fully exclude a combination of dragging effects mixed with DR. Thus, as
an additional quantitative test we have converted the measured $\alpha_{\rm
mb}(\xi = 40)$ to a model of the rate coefficient expected at the lower
expansion factor $\alpha^{''}_{\rm mb}(\xi = 20)$. A real DR resonance, as
opposed to a dragging effect, should appear similar in both the converted
$\alpha^{''}_{\rm mb}(\xi = 20)$ and the measured $\alpha_{\rm mb}(\xi = 20)$.
To generate $\alpha^{''}_{\rm mb}(\xi = 20)$ we deconvolved the $\alpha_{\rm
mb}(\xi = 40)$ data using the corresponding energy distribution $f_{\rm
mb}(\xi=40)$ and the fitting procedure described in Section
\ref{sec:crossecdernew}. We then reconvolved the extracted cross section with
$f_{\rm mb}(\xi=20)$ to obtain the model of experimental rate coefficient
$\alpha_{\rm mb}^{''}(\xi = 20)$. 

Figure \ref{fig:fexp40} compares the measured rate coefficient $\alpha_{\rm
mb}(\xi = 20)$ and the converted $\alpha_{\rm mb}^{''}(\xi = 20)$. The measured
$\alpha_{\rm mb}(\xi = 40)$ and the corresponding model $\alpha'_{\rm mb}(\xi =
40)$ are also displayed. We find satisfying agreement between $\alpha_{\rm
mb}(\xi = 20)$ and $\alpha^{''}_{\rm mb}(\xi = 20)$. The small differences
remaining are attributed, in part, to uncertainties in the scaling used to
derive $\ktperp$ for $\xi=40$ from that at $\xi=20$. To first order we expect
$\ktperp$ to scale as $\xi^{-1}$, but there may be small nonlinear terms which
we have ignored. These may introduce uncertainties in the $f_{\rm mb}(\xi=40)$
used which could carry over into the extracted cross section and finally into
$\alpha^{''}_{\rm mb}(\xi = 20)$. Additionally, we did not measure $\alpha_{\rm
mb}(\xi = 40)$ at energies of $\Ed>20$~meV. So, above this point we have used
the $\alpha_{\rm mb}(\xi = 20)$ data. This may produce small discrepancies at
energies $\Ed\gtrsim10$~meV.

\subsection{Recombination Cross Section}\label{sec:rescrosssec}
We have converted the experimental DR rate coefficient $\alpha_{\rm
mb}(\xi=20)$ to a cross section using a procedure introduced in Section
\ref{sec:crossecdernew} and described in detail in Appendix \ref{app:A}.
The resulting $\sigma'(E)$ is plotted in Figure~\ref{fig:res_crosssec}
for the energy range $E=0-4.5$~eV. The lower edge of the first energy bin is set
to $E=0$ and is not displayed on the logarithmic energy scale of the plot.

Deriving the cross section  involves fitting  $\alpha_{\rm mb}$ with a sum of
model rate coefficients, each convolved using a single cross section bin. These
sub-functions, $\sigma'_i\,\Psi_i(\Ed)$, are plotted in Figure
\ref{fig:rateexpA}. We plot the corresponding collision energy distribution
function $f_{\rm mb}$ for several values of $\Ed$ in Figure~\ref{fig:fmb}. To
ensure physical meaning of the results we limited the fitting parameters
$\sigma'_i$ to positive values only. The fitting yielded a minimum
$\chi^2/N_{\rm DF} = 1.02$, where $N_{\rm DF} = 30$ is the number of degrees of
freedom in the fit. 

To explore the numerical stability of the fitting, we have derived the cross
section from a set of 1000 simulated merged beams spectra $\alpha_{\rm mb}^{\rm
sim}$.  Each data point in a given simulated spectrum was obtained from the sum
of the experimental value $\alpha_{\rm mb}(\Ed)$ and an additive shift which was
randomly chosen from a Gaussian distribution centered around zero and with a
width equal to the 1$\sigma$ statistical error in $\alpha_{\rm mb}(\Ed)$. Each
$\alpha_{\rm mb}^{\rm sim}$ was then fitted to derive a cross section. The cross
section values in each energy bin nearly follow a normal distribution. In some
bins, however, up to 8\% of the fitted $\sigma_i$ values converged to zero. We
attribute this behavior to the numerical instability for some of the fits,
possibly due to converging to a local $\chi^2$ minimum or  because of strong
correlations with cross section values in neighboring energy  bins. The mean
cross section values for each bin are displayed  in Figure
\ref{fig:res_crosssec}. The standard deviations above and below the mean are
displayed by asymmetric vertical error bars and reflect the statistical errors
propagated from $\alpha_{\rm mb}$. The mean cross section values differ from
those obtained directly from the experimental data by only a small fraction of
the standard deviations. To test the accuracy of the mean cross section we have
convolved it using equation (\ref{eq:ratecross}) to generate a merged beams rate
coefficient and compared the result to $\alpha'_{\rm mb}$, which was derived by
fitting the experimental data. The difference between the two is less than a
fifth of the statistical uncertainties in our experimental data. 

Additionally we have investigated the sensitivity of the cross section to the
uncertainties in $\ktperp$ and $\ktpar$. We have used their extreme values
$\ktperp = 1.3-2.0$~meV and $\ktpar=20-70~\mu$eV to generate $f_{\rm mb}$ and
derived the cross section from the 1000 model rate coefficients, as discussed
above. The mean cross sections were then compared to the one obtained with the
most probable electron beam temperatures $\ktperp = 1.65$~meV and
$\ktpar=25~\mu$eV. The differences originating from $\ktperp$ and $\ktpar$ were
added in quadrature and are displayed by gray bars in
Figure~\ref{fig:res_crosssec}. The sensitivity of $\sigma'(E)$ to the electron
beam parameters generally decreases with increasing energy. This is due to the
decreasing relative energy spread $\Delta E/E$ with increasing energy. There is,
though, an enhancement of the sensitivity seen in few bins at $E\sim0.03$~eV.
We attribute this to narrow structures in the cross section with spacing
comparable to the energy resolution at these energies (of order 2.5~meV).
The effect of the electron beam temperature parameters is significantly lower
than the statistical errors at all energies.  Lastly, the 12\% uncertainty on
the absolute scaling of $\alpha_{\rm mb}$ directly propagates to $\sigma'$.
However, this error is too small to be seen in Figure
\ref{fig:res_crosssec}.

\subsection{Plasma Recombination Rate Coefficient}\label{sec:resplasma}
\begin{figure}
    \plotone{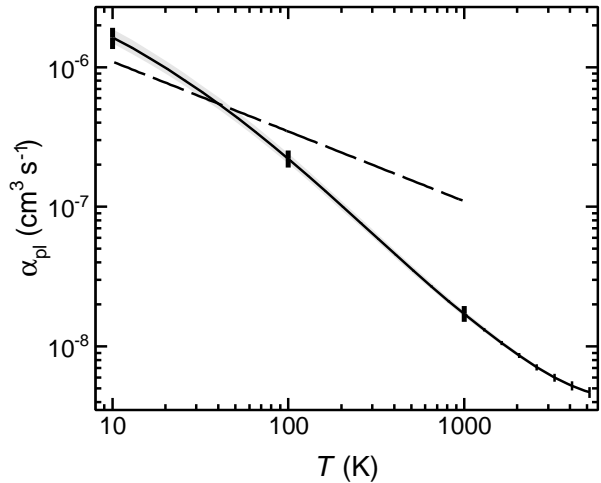}
    \caption{\doublespace
The experimentally derived DR plasma rate coefficient for \HClp is
shown by the black line. The thin error bars display the statistical
uncertainties in $\alpha_{\rm mb}$ propagated through our method to generate a
plasma rate coefficient. These error bars are only visible above $\sim 1000$~K.
The thick error bars mark the total systematic uncertainty originating from the
error on the absolute scaling and from the error due to uncertainties on
$\ktperp$ and $\ktpar$. The gray area shows the error originating solely from
uncertainties in $\ktperp$ and $\ktpar$. The dashed line shows $\alpha_{\rm
pl}^{\rm di}$, the rate coefficient typically assumed for diatomic molecular
ions. The internal excitation temperature of the \HClp ions in this experiment
lies near 300 K.
	\label{fig:rateplasmac2}
	}
\end{figure}

We have used the model cross section $\sigma'$ as a representation of the DR
cross section $\sigma$ and converted it to a plasma rate coefficient using the
procedures described in Section \ref{sec:plasmader}. The resulting $\alpha_{\rm
pl}(T)$ is plotted in Figure~\ref{fig:rateplasmac2} for a plasma temperature
range $T=10-5000$~K. 

To propagate the statistical uncertainty from the measured $\alpha_{\rm
mb}$ to the derived $\alpha_{\rm pl}$ we follow the technique used for the cross
section in Section \ref{sec:rescrosssec}. First we created a set of 1000
simulated merged beams spectra $\alpha_{\rm mb}^{\rm sim}$ based on the measured
$\alpha_{\rm mb}$ and the corresponding statistical uncertainty. Each
$\alpha_{\rm mb}^{\rm sim}$ was used to derive a cross section which was
subsequently convolved with $vf_{\rm pl}$ to generate a plasma rate coefficient
$\alpha_{\rm pl}^{\rm sim}$. The spread in the values of $\alpha_{\rm pl}^{\rm
sim}$ at each $T$ closely followed a normal distribution. The Gaussian width of
this spread at each $T$ was taken as the $1\sigma$ statistical accuracy for
$\alpha_{\rm pl}$. For $T$ below 400~K these errors are less than 1\%. They
increase at higher temperatures and
reach $8\%$ at 5000~K. 

The uncertainties in $\ktperp$ and $\ktpar$ introduce an error in the derived
$\alpha_{\rm pl}$. The plasma rate coefficient displayed in Figure
\ref{fig:rateplasmac2} was derived for $\ktperp=1.65$~meV. We have also
evaluated $\alpha_{\rm pl}$ taking into account the $\pm0.35$~meV uncertainty in
$\ktperp$ and repeated the analysis using the limiting values of $\ktperp$. The
corresponding respective change of $\alpha_{\rm pl}$ is $\pm12$\% at $T=10$~K,
decreases to $\pm8$\% at 100~K, and is less than $\pm5$\% above 1000~K. Varying
$\ktpar$ within the estimated range of values has a much smaller effect. At
$T=10$~K the change in $\alpha_{\rm pl}$ is $^{+3.5}_{-0.5}$\% and drops to less
than $^{+0.5}_{-0.1}$\% at $T\geq50~K$. In this paragraph positive (negative)
changes in $\alpha_{\rm pl}$ correspond to increased (decreased) $k_{\rm B}T_\|$
and $\ktpar$.

The only other important systematic uncertainty in $\alpha_{\rm pl}$ is the 12\%
absolute scaling error propagated from $\alpha_{\rm mb}$. We treat this scaling
error and those from $\ktperp$ and $\ktpar$ as independent and add them in
quadrature. The resulting total systematic error amounts to 17\% at 10~K, 14\%
at 100~K, and less than 13\% above 1000~K.

We have fitted our results for ease of use in astrochemical models.  
We found that neither the two-parameter function commonly used to describe DR
plasma rate coefficients (e.g., \citealt{Florescu:PR:2006}) nor the more
general three-parameter extension used in astrochemical databases (e.g.,
\citealt{UMIST2006, Wakelam:ApJS:2012}) are able to fit our measured plasma rate
coefficient accurately over the entire temperature range from 10 to 5000~K. Such
fits do not reproduce our results to within better than 40\%. Therefore we
propose a modified form of the two-parameter function, namely
\begin{equation}
\alpha_{\rm pl}^{\rm fit}(T) =
A\,\left(\frac{300}{T}\right)^n + B ,
\label{eq:plasmafitnew}
\end{equation}
where
\begin{equation}
B = T^{-3/2}\sum_{i=1}^4 c_i \exp(-T_i/T)
\label{eq:B}
\end{equation}
This new function allows for more accurate fits than the previously used fitting
functions. The results of fitting equation (\ref{eq:plasmafitnew}) to our data
over the full temperature range are given in Table~\ref{tab:plasmares}.
The deviations of $\alpha_{\rm pl}^{\rm fit}$ from the data are less than 1\%
over the full temperature range. It should be emphasized that the internal
temperature of the \HClp ions in the experiment yielding these result is
expected to lie near 300~K.

\section{Discussion}
\label{sec:discuss}
\subsection{Comparison with previous analysis methods}
The principal difference between our new method for deriving the cross section
and the plasma rate coefficient compared to the methods used in previous merged
beams DR measurements lies in the conversion of the merged beams rate
coefficient to a cross section. Our new method is essentially a one-step
process. The traditional method, though, requires two steps. First it corrects
the merged beams rate coefficient for the increased electron-ion collision
energy in the merging and de-merging regions while ignoring the electron
velocity spread \citep{Lampert:PRA:1996}. Then the corrected rate coefficient is
deconvolved to yield a cross section, assuming parallel beams in the interaction
zone and zero electron energy spread along the beam axis, i.e., $\ktpar=0$
\citep{Mowat:PRL:1995}. Our approach allows us to estimate the errors arising
from the assumptions for each of these two steps.

To test the validity of the \cite{Lampert:PRA:1996} step, we have extracted the
cross section from $\alpha_{\rm mb}$ using our new approach and converted it to
$\alpha^*_{\rm mb}$ using equation (\ref{eq:ratecrosscorr}). The resulting
$\alpha^*_{\rm mb}$ then represents the rate coefficient as it would be measured
in a merged beams configuration with parallel beams only and no merging or
de-merging sections. Next we have reanalyzed our measured $\alpha_{\rm mb}$ data
using the method of \cite{Lampert:PRA:1996} to generate $\alpha^{* \rm L}_{\rm
mb}$. Comparing the two data sets $\alpha_{\rm mb}^*$ and $\alpha_{\rm mb}^{*
\rm L}$ at high energies, where $\Ed \gtrsim \ktperp$, our correction procedure
and the one proposed by \cite{Lampert:PRA:1996} are equal to within their
statistical accuracies. However, at lower energies, the traditional correction
results in a merged beams rate coefficient which is up to $\sim5$\% higher.
Both the toroid correction methods of \citeauthor{Lampert:PRA:1996} and ours
increase the merged beams rate coefficient at
these low energies. Therefore the difference can be interpreted as an
over-correction by the older approach. We attribute this difference to
the neglected electron velocity spread in the \citeauthor{Lampert:PRA:1996}
method.

Next we tested the combined effects of the \cite{Lampert:PRA:1996}  and the
\cite{Mowat:PRL:1995} steps. For this we employed our new method for the cross
section derivation with two modifications. First, we used $\alpha^{* \rm L}_{\rm
mb}$ as the input rate coefficient, and second, we generated the collision
energy distribution for straight beams $f_{\rm mb}^*$ while setting $\ktpar=0$.
This latter step simulates the \citeauthor{Mowat:PRL:1995}\ approximation. We
then converted the extracted cross section to the plasma rate coefficient
$\alpha^{\rm M}_{\rm pl}$ using equation (\ref{eq:rateplas}).  The resulting
plasma rate coefficient is larger than $\alpha_{\rm pl}$ by $2.3$\% at $T=10$~K,
$4.2$\% at $T=100$~K, and to $1.7$\% at $T=1000$~K. The spread in these values
due to the uncertainties in $\ktpar$ and $\ktperp$ is less than $\pm0.9\%$.

The errors in $\alpha_{\rm pl}$ for \HClp DR which are introduced by the
\cite{Lampert:PRA:1996} and \cite{Mowat:PRL:1995} approximations are only a few
percent, much smaller than the other systematic errors in the measurement.
However, larger differences could potentially arise if the measured merged beams
rate coefficient were  highly structured or decreased more rapidly with
increasing energy. Additionally, it is not trivial to implement a Fast Fourier
Transform (FFT) method, such as recommended by \cite{Mowat:PRL:1995}, using DR
data on a non-uniform energy grid. Furthermore, the FFT method does not weigh
the input data points by their statistical importance. All of these issues are
readily accounted for by our new approach. The advantages of our new method are
likely to be critical for reliably analyzing the expected highly structured DR
data from upcoming Cryogenic Storage Ring (CSR) experiments on rotationally cold
molecular ions \citep{Von_hahn:NIM:2011,Krantz:JPCS:2011}.

\subsection{Experimental DR rate coefficient}
Up to $\Ed \approx 0.5$~eV, the \HClp DR rate coefficient $\alpha_{\rm mb}$
decreases rapidly with increasing energy. This behavior is typical for low
energy DR spectra \citep{Florescu:PR:2006, Larsson:book:2008}. Yet, as shown in
Figure \ref{fig:rateexpA}, the slope between $\Ed = 4$ and 30~meV is much
steeper than the rate coefficient derived from the $\sigma \propto E^{-1}$
expected for the direct DR process \citep{Bates:PR:1950,Larsson:book:2008}. As
described in Section~\ref{sec:DRpaths}, direct electron capture to a repulsive
neutral potential surface is unlikely at these low collision energies. Thus at
these energies, DR is probably dominated by the indirect process and the
enhanced rate coefficient most likely results from numerous DR resonances. These
are unresolved due to both the energy resolution of the experiment and the
natural widths of the resonances.

Based on the collision energies these resonances probably originate from neutral
Rydberg states converging to a rotationally excited \HClp ${\rm X}\,^2 \Pi$
core. The exact positions for the resonances cannot be easily determined because
of the large number of combinations between the initial $\sim 300$~K
distribution of \HClp excited states and the many energetically accessible
ro-vibrationally excited levels in the neutral Rydberg system. Still, the steep
decrease of $\alpha_{\rm mb}$ between $\Ed=4-30$~meV may be attributed to the
rotational structure of \HClp. As discussed in Section~\ref{sec:DRpaths}, each
rotational level of the ion is expected to be the end point for a Rydberg DR
resonance series. In Figure~\ref{fig:rateexpA} we plot the range of such end
point energies allowing for $\Delta J = 1$, 2, and 3 changes of the angular
momentum with respect to the initial ionic state. The range of the end points
(indicated by the gray bars) is due to the initial rotational excitation of the
\HClp ions (using rotational parameters from \citealt{Sheasley:JMS:1973}; the
levels spacings are nearly the same for ions in the ${\rm X}\,^2 \Pi_{3/2}$ or
${\rm X}\,^2 \Pi_{1/2}$ state, respectively.) The energy range of the fastest
decrease in $\alpha_{\rm mb}$ matches best the lowest possible change of the
angular momentum, i.e., $\Delta J = 1$. Thus, it appears likely that indirect
processes involving rotational excitation of the $\sim 300$~K \HClp ions play an
important role for the DR rate at the lowest energies.  Correspondingly,
rotational excitation rates by low energy electron collisions of \HClp may be
large.

Somewhat more resolved structures appear between 30 and 300~meV. There is the
suggestion of a neutral Rydberg series limit at $\sim 80$~meV corresponding to
the $3/2\rightarrow1/2$ \HClp fine-structure transition
\citep{Sheasley:JMS:1973} and at $\sim 0.3$~eV corresponding to the
$v=0\rightarrow 1$ excitation \citep{NISTWebBookWhole}. The structures are
blurred partly by the energy resolution of the experiment and partly by the
initial rotational distribution of the stored \HClp ions. Further interpretation
cannot be given without more detailed calculations.

The increase in $\alpha_{\rm mb}$ at $\Ed\gtrsim0.6$~eV can be attributed to
opening of additional dissociation pathways forming either excited Cl (above
0.6~eV) or excited H (above 1.9~eV) or both. Moreover, Rydberg resonances
involving the A$^2\Sigma^+$ electronically excited \HClp core may also be
present. Ion pair production, which can reduce the DR signal as discussed in
Section~\ref{sec:DRpaths}, will set in above 1.7~eV. At these energies, the
statistical uncertainties in our data prevent us from being able to discern the
impact on the DR signal due to the opening up of these channels. DE channels are
not accessible in the investigated collision energy range.

\subsection{Cross Section}
The derivation of the DR cross section from the merged beams rate coefficient
makes the DR data independent of the experimental configuration. On the other
hand, the experimental energy spread together with the statistical quality of
the rate coefficient data limits the resolving power of our method for deriving
the cross section.  We have adjusted the cross section binning such that
it provides good energy resolution while keeping the numerical instabilities
small. The interpretation of the derived cross section is therefore somewhat
limited and one needs to proceed cautiously in interpreting any structures which
have an energy width comparable to the energy binning. But given that caveat,
much of the structure discussed in above in $\alpha_{\rm mb}$ can also be seen
in the cross section.

\subsection{Plasma rate coefficient}
Our experimentally derived DR plasma rate coefficient for \HClp displays an
unusually steep slope. Existing astrochemical models usually approximate DR of
\HClp by \citep{Neufeld:ApJ:2009} $$\alpha_{\rm pl}^{\rm
di}\approx2.0\e{-7}\times(300/T)^{0.5}~{\rm cm}^3\,{\rm s}^{-1}.$$ This is
believed to correspond to DR of typical diatomic molecular ions
\citep{Florescu:PR:2006}. However, some astrochemical databases use a value 1.5
times higher for \HClp (e.g., \citealt{UMIST2006}), though there is no obvious
reason for this. Taking the ratio of $\alpha_{\rm pl}^{\rm}/\alpha_{\rm pl}^{\rm
di}$ yields 1.5, 1.1, 0.64, 0.33, and 0.16 at $T=10$, 30, 100, 300, and 1000~K,
respectively. Such differences are larger than our experimental error bars. Thus
we find that the ``typical'' diatomic DR rate coefficient is incorrect in both
magnitude and temperature dependence for HCl$^+$. Note, however, that our
derived rate coefficient is for an internal temperature of $\sim$300~K of the
\HClp ions.

\subsection{Astrochemical implications}
Our new data suggest that HCl$^+$ depletion by DR in the cold ISM ($\sim
10-50$~K) is faster than or similar to that currently assumed in existing
astrochemical models. However, at higher temperatures the new data display a
slower HCl$^+$ destruction by DR. As HCl$^+$ leads to the formation of
H$_2$Cl$^+$, the H$_2$Cl$^+$ abundance will also decrease or increase when using
our new data, depending on the temperature. Unfortunately, the kinetic
temperatures of the observed environments were not directly derived in the works
of \cite{Lis:AA:2010}, \cite{de_luca:ApJL:2012}, and \cite{Neufeld:ApJ:2012}.
Model calculations \citep{Neufeld:ApJ:2009} predict that there is an abundance
maximum of \HClp in the outer parts of dark interstellar clouds which, at an
opacity of $A_v \sim 1$, are less dense than the core and with a temperature of
$\sim 500$K~ are also hotter. In these regions DR is predicted to clearly
dominate the destruction of \HClp.  Thus, our measured DR rate coefficient
(which is about a factor of 4 lower than the ones used in current astrochemical
models) will augment the predicted \HClp abundances  in that region, leading to
an improved agreement with the  observed \HClp densities. It will also affect
the H$_2$Cl$^+$ abundance in the outer parts of the clouds (which also predicted
an abundance maximum at roughly the same cloud density as  the one for \HClp is
found), since a lesser efficient competition by the DR of \HClp will allow more
H$_2$Cl$^+$ to be formed by reaction of \HClp with H$_2$. 

\cite{Neufeld:ApJ:2012} have explored the effect of lowering the \HClp DR rate
coefficient on the discrepancy between the predicted and observed \HClp and
H$_2$Cl$^+$ abundances. In their astrochemical model they decreased $\alpha_{\rm
pl}$ of \HClp to $\alpha_{\rm pl}^{\rm di}/10$. The resulting abundances of
\HClp and H$_2$Cl$^+$ increased by factors of 2.7 and 1.3, respectively. In both
cases the calculated abundances are still far too low to match the greater than
ten times larger observed abundances. The result of their tests suggests that
our new \HClp DR data, which are about four times smaller than  $\alpha_{\rm
pl}^{\rm di}$ at the relevant temperature, cannot fully explain the discrepancy
between current astrochemical models and the observations. Still, 
our new data differ from $\alpha_{\rm pl}^{\rm di}$  not only in magnitude, but
also in the  temperature dependence. Therefore the test of
\cite{Neufeld:ApJ:2012} does not directly describe the effect of
our new DR data and their calculations should be repeated in order to
fully evaluate the remaining discrepancy.

We also note that our experimental data are for HCl$^+$ rotationally excited to
$T_{\rm rot}\sim300$~K. In the cold ISM, however, the internal excitation is
expected to be much lower. In the very low density environment in interstellar
clouds, the collision rate is much lower than the typical radiative decay time
\citep{Spitzer:ISM:1978}. Thus $T_{\rm rot}$ is generally even lower than the
kinetic temperature $T$ and most of the molecules are expected to be in their
rotational ground state. The response of the DR rate coefficients to $T_{\rm
rot}$ has not yet been studied systematically. The few existing studies on light
ions find that the measured merged beams rate coefficient exhibits an
increasingly rich resonant structures with decreasing $T_{\rm rot}$ (e.g.,
\citealt{Amitay:PRA:1996}, \citealt{Zhaunerchyk:PRL:2007},
\citealt{Petrignani:PRA:2011}, \citealt{Schwalm:JPCS:2011}). This can be
explained by the smaller number of rotational states that contribute with
decreasing $T_{\rm rot}$.  The individual resonant structures thereby become
resolved as they no longer overlap with resonances of other, energetically
higher states. Low values of $T_{\rm rot}$ can result in either a lower or
higher plasma rate coefficient, depending on the particular shape of the cross
section. We expect that future DR measurements at the CSR facility will be able
to address this issue by investigating molecular ions rotationally colder than
what can be achieved with TSR.

\section{Summary}\label{sec:sum}
We have measured the absolute DR rate coefficient for \HClp in a merged beams
configuration at electron-ion collision energies from 0~eV to 4.5~eV. Using a
novel method, we have converted the experimental merged beams rate coefficient
to a cross section and then to a plasma rate coefficient. Our new approach
provides more precise results and better control on uncertainties as compared to
previously used methods. At molecular cloud temperatures below $\sim 50$~K, the
resulting plasma rate coefficient is similar or faster than that typically
assumed for DR of other diatomic molecular ions. However, at higher temperatures
relevant for HCl$^+$ observations, DR of HCl$^+$ is slower than previously
expected. Thus our data indicate that the current issues in modeling chlorine
chemistry in interstellar clouds may be partly resolved by our new DR data.

\acknowledgments
We thank the MPIK accelerator and TSR crews for their excellent support. ON and
DWS were supported in part by the NSF Division of Astronomical Sciences
Astronomy and Astrophysics Grants program and by the NASA Astronomy and Physics
Research and Analysis Program. DS acknowledges support by the
Weizmann Institute of Science through the Joseph Meyerhoff program. The work is
supported in part by the German-Israeli Foundation for Scientific Research (GIF
under contract nr. I-900-231.7/2005). WG acknowledges partial support by the
COST Action CM0805: ``The Chemical Cosmos: Understanding Chemistry in
Astronomical Environments''. BY thanks for the support from MPG and from CAS-MPS
program. The work is supported in part by the DFG Priority Program 1573
``Physics of the Interstellar Medium" and by the Max Planck Society. 

\appendix
\section{Deconvolving the measured merged beams rate coefficient}\label{app:A}
In Section \ref{sec:crossecdernew} we briefly introduced a new method for
deriving a cross section from the measured merged beams rate coefficient. This
approach begins with a model cross section $\sigma'(E)$ which we multiply by the
relative collision velocity and then, following equation (\ref{eq:ratecross}),
convolve the product with the center of mass energy distribution function
$f_{\rm mb}(E,\Ed,T_{||},T_\perp,\boldsymbol{X})$.  This generates a model rate
coefficient $\alpha'_{\rm mb}(\Ed)$. We iteratively modify the parameters
defining the shape of $\sigma'(E)$ to minimize the $\chi^2$ between the measured
$\alpha_{\rm mb}(\Ed)$ and the model $\alpha'_{\rm mb}(\Ed)$. In this Appendix
we discuss the core of our deconvolution method, namely the numerical evaluation
of equation (\ref{eq:ratecross}).

The required collision energy distribution $f_{\rm mb}(E,\Ed,T_{||},
T_\perp,\boldsymbol{X})$~is not known in analytical form.  So we use a Monte
Carlo simulation which includes the effects of both the electron velocity spread
and the beam overlap geometry. The Monte Carlo method requires a large number of
simulated events to reach a low statistical uncertainty for the multidimensional
$f_{\rm mb}$. The resulting computational demands rise steeply as the power of
the number of dimensions. This can be partly reduced by fixing the electron
energy spreads $\ktpar$ and $\ktperp$ and also fixing the merged beams geometry
$\boldsymbol{X}$\!, as none of these parameters change during the measurement
and can thus be kept constant during the fit. 

The required computational time can be further reduced by using a good
discretization of the $E$ dimension. We model $\sigma'(E)$ as a histogram with
non-uniform bin widths. For $E\lesssim\ktperp$, we choose bin widths similar to
the experimental energy resolution for $E=\Ed$. At higher energies we use
broader bins, but which are still much narrower than the structure typically
observed in TSR DR measurements of $\alpha_{\rm mb}(\Ed)$. A consequence and
also advantage of this approach is that the exact energy bin positions and
widths are not critical and can be fixed during fitting without any significant
loss in the precision of the fit.

With a fixed energy binning we can then transform the model form of  equation
(\ref{eq:ratecross}) into the summation
\begin{equation}
\alpha'_{\rm mb}(\Ed) = \sum_i \sigma'_i \int_{E_i}^{E_{i+1}} \,v\,f_{\rm
mb}(E, \Ed, T_{||}, T_\perp, \boldsymbol{X})\,dE,
\label{eq:ratecrosssum}
\end{equation}
where $\sigma'_i$ is the model cross section in $i^{\rm th}$ energy bin and
$E_i$ and $E_{i+1}$ are the edges of the bin. 
Defining the function
\begin{equation}
\Psi_i(\Ed, T_{||}, T_\perp, \boldsymbol{X}) \equiv \int_{E_i}^{E_{i+1}}
\,v\,f_{\rm mb}(E, \Ed, T_{||}, T_\perp, \boldsymbol{X})\,dE,
\label{eq:psi}
\end{equation}
we can re-express equation (\ref{eq:ratecrosssum}) as
\begin{equation}
\alpha^\prime_{\rm mb}(\Ed)
= \sum_i \sigma'_i\,\Psi_i(\Ed, T_{||}, T_\perp, \boldsymbol{X}).
\label{eq:lincomb}
\end{equation}
Because the $\Psi_i$ integrals are independent of the fitting parameters
$\sigma'_i$, the integrals can be calculated prior to the iterative
$\chi^2$-minimization procedure. As a result, the overall fitting time is
reduced and is dominated by the initial generation of the $\Psi_i$ integrals. 
For a given required statistical accuracy, this scales linearly with number of
cross section bins $\sigma'_i$  and with the number of fitted energy points
$\Ed$. Additionally, because the form of equation (\ref{eq:lincomb}) is a linear
combination of the constant factors $\Psi_i$ scaled by the fitting parameters
$\sigma'_i$, it allows us to use efficient fitting algorithms (e.g.,
\citealt{Press:2007:Cambridge}).

We calculate the $\Psi_i$ integrals employing a numerical Monte Carlo method
(e.g., \citealt{Press:2007:Cambridge}). In short, we use a model of the merged
beams geometry in which we generate a large number of collision events according
to the fixed experimental parameters $\Ed$, $T_{||}$, $T_\perp$, and
$\boldsymbol{X}$. Events are generated by randomly selecting the dissociation
position in $\boldsymbol{X}$ as well as the electron velocity vector for the
given values of $\Ed$, $T_{||}$ and $T_\perp$.  The exact
probability distributions used are discussed below. 

\begin{figure}
    \plotone{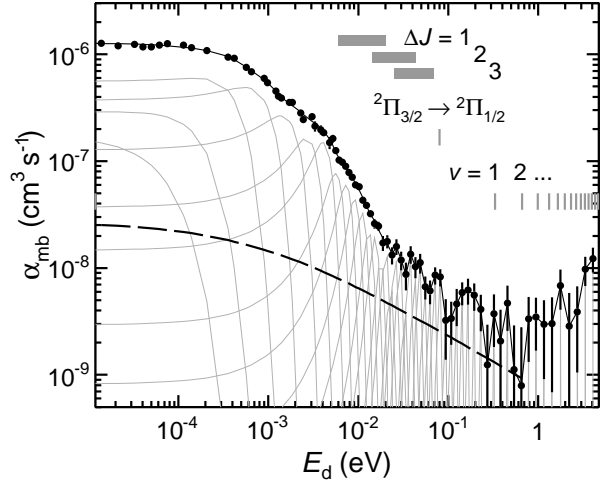}
    \caption{\doublespace
The experimental merged beams rate coefficient $\alpha_{\rm mb}$ for DR of \HClp
acquired with an adiabatic magnetic expansion factor of $\xi=20$. The full
circles show the data and the error bars indicate the 1$\sigma$ statistical
confidence level. Some of the error bars are smaller than the diameter of the
plotted data points. The long dashed line illustrates the shape of the merged
beams rate coefficient expected for direct DR, i.e., for $\sigma(E)\propto
E^{-1}$, and is arbitrarily scaled on the vertical axis. To determine the cross
section $\sigma'$, the model rate coefficient $\alpha'_{\rm mb}$ (black full
line) was fitted to $\alpha_{\rm mb}$. The gray curves indicate the best fit
sub-functions $\sigma'_i\,\Psi_i$, each corresponding to an individual convolved
cross section bin. See Appendix \ref{app:A} for more details. The gray bars mark
the range of rotational excitation thresholds with changes of angular momentum
$\Delta J=1$, 2, and 3 for initial levels from $J=3/2$ to $J=15/2$, i.e., the
levels dominantly populated at 300~K. The fine structure transition energy and
the vibrational excitation thresholds are labeled by vertical gray lines. All
energies are given with respect to the \HClp(${\rm X}\,^2\Pi_{3/2}$) ground
state.
\label{fig:rateexpA}
	}
\end{figure}

\begin{figure}
    \plotone{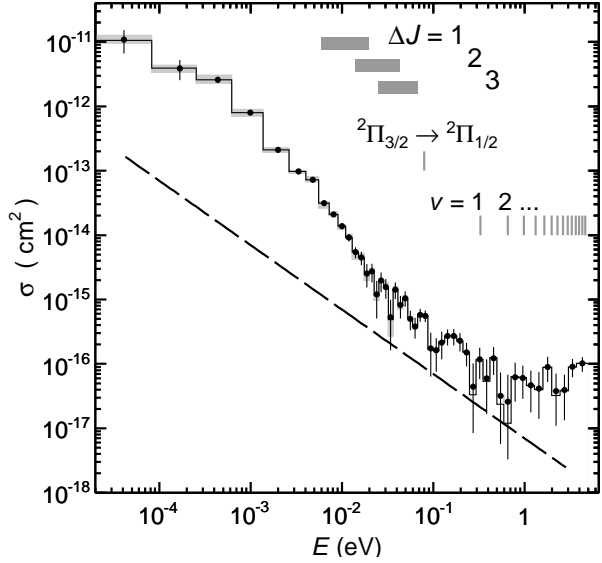}
    \caption{\doublespace
	\label{fig:res_crosssec}
	The DR cross section for \HClp is shown by the full line histogram.
The non-displayed lower edge of the left-most energy bin is at $E=0$. The
vertical error bars display the standard deviation of cross section values
obtained from deconvolving 1000 simulated merged beams rate coefficients. A
full explanation is given in Section \ref{sec:rescrosssec}. The gray bars
along the data show the error originating from uncertainties in $\ktperp$ and
$\ktpar$. The long-dashed line illustrates the
shape of merged beams rate coefficient expected for a direct DR process, i.e.,
$\sigma(E)\propto E^{-1}$. The curve is arbitrarily scaled on the vertical axis.
The rotational, fine-structure, and vibrational thresholds
are the same as in Figure~\ref{fig:rateexpA}.
}
\end{figure}

For each event $j$, we calculate the electron-ion collision velocity $v_j$ and
the corresponding center of mass collision energy $E_j$. The propagation through
the model of the randomized dissociation position and electron velocity vector
ensures that the set of values generated for $E_j$ follows the desired $f_{\rm
mb}(E)$ distribution. Hence the numerical integration of equation (\ref{eq:psi})
reduces to a simple summation given by
\begin{equation}
\Psi_i(\Ed, T_{||}, T_\perp, \boldsymbol{X}) = \frac{1}{N_{\rm s}}
\sum_{j=1}^{N_{\rm s}} \left\{
\begin{array}{ll} 
v_j, & E_i\leq E_j < E_{i+1} \\
0 , & {\rm otherwise}
\end{array} \right. ,
\label{eq:psi2}
\end{equation}
where $N_{\rm s}$ is the total number of simulated events. For each energy
$\Ed$, we generate $N_{\rm s} = 10^7$ events.   This provides a statistical
precision better than $\sim0.1\%$ for the resulting $\alpha^\prime_{\rm
mb}(\Ed)$. The fitting sub-functions $\sigma_i\Psi_i(\Ed)$  are shown in
Figure~\ref{fig:rateexpA}.

\begin{figure}
    \plotone{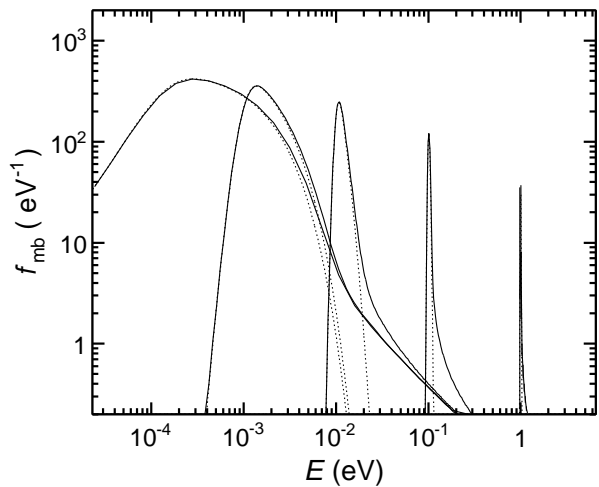}
    \caption{\doublespace
Simulated energy distribution functions $f_{\rm mb}$ (full line) for
$\Ed=10^{-4}$~eV, $10^{-3}$~eV, $10^{-2}$~eV, $10^{-1}$~eV, and 1~eV (from left
to right) for electron beam parameters $\ktperp=1.65$~meV, $\ktpar=25~\mu$eV,
and the electron beam geometry of the Target. For comparison we plot the energy
distribution functions for parallel beams $f_{\rm mb}^*$ (dotted line). All
$f_{\rm mb}^*$ functions are scaled so that the maxima
match those of corresponding $f_{\rm mb}$ functions.
	\label{fig:fmb}
	}
\end{figure}

Although not directly needed for the cross section derivation, one can similarly
obtain an average value of the distribution function
$f_{\rm mb}$ in energy bin $i$ as 
\begin{equation}
f_{{\rm mb},i}(\Ed, T_{||}, T_\perp, \boldsymbol{X}) = \frac{1}{N_{\rm s}}
\sum_{j=1}^{N_{\rm s}} \left\{
\begin{array}{ll} 
1, & E_i\leq E_j < E_{i+1} \\
0 , & {\rm otherwise}
\end{array} \right. .
\label{eq:fmbsim}
\end{equation}
The histograms $f_{{\rm mb},i}$ are plotted in Figure~\ref{fig:fmb} for
several values of $\Ed$. Here we have chosen a very narrow energy binning so
that the histogram $f_{{\rm mb},i}$ closely approximates $f_{{\rm mb}}$.
For comparison, we also plot in Figure~\ref{fig:fmb}, the energy distribution
function for parallel beams $f_{\rm mb}^*$.  This helps to highlight the
extension of the high energy tail $f_{\rm mb}$ due to the merging and demerging
of the beams.

We model the electron beam as a cylindrical body with the shape described in
Section \ref{sec:setup}. Given that the ion beam diameter is much smaller than
the electron beam and that the magnetic fields in the Target are too weak to
significantly affect the ions, we can approximate the ion beam as a straight
line of zero diameter. We assume uniform electron and ion beam densities within
the model beams and can therefore treat the event probability distribution as
uniform over the entire interaction region.

Because of the large difference between the electron and ion masses and the
phase-space cooling, the velocity spread of the cooled ion beam has a negligible
effect on the experimental energy spread $f_{\rm mb}$. Thus we can model the
ions with a constant laboratory velocity of $v_{\rm i} = \sqrt{2E_{\rm
cool}/m_{\rm e}}$ with no velocity spread. The electron velocity vectors are
randomized in the laboratory frame. The orientation of the electron beam bulk
velocity is determined from the electron beam geometry at the position of each
simulated event. The velocity components perpendicular to the electron beam axis
are described by a Gaussian distribution centered around $v_{\rm e,\perp} = 0$
and with a width of $\sigma_\perp =\sqrt{k_{\rm B}T_\perp/m_{\rm e}}$. In the
parallel direction the velocity distribution is centered at $v_{\rm e,\|} =
v_{\rm i} + \sqrt{2\Ed/m_{\rm e}}$ with a Gaussian width of $\sigma_\|
=\sqrt{k_{\rm B}T_{||}/m_{\rm e}}$. The electron-ion collision velocity $v_j$ is
obtained by subtracting the ion and electron velocity vectors. The corresponding
energy $E_j$ is then used for the summation condition in equation
(\ref{eq:psi2}) to decide to which $\Psi_i$ each event $j$ contributes.

Lastly we note that choosing an appropriate energy binning for the model
cross section is essential in order to avoid numerical instabilities when
fitting the model $\alpha'_{\rm mb}$ to the measured $\alpha_{\rm mb}$.
Bins significantly narrower than the energy resolution or than the energy
spacing between the measured data points give an under-determined
fitting system which results in fitted $\sigma'_i$ values with large
uncertainties strongly correlated with neighboring bins. On the other hand,
using bins which are too broad results in too coarse of a fitting function and
a poorer quality of the fit. 

We have developed a general recipe to define an optimal cross section energy
binning which does not suffer from these effects. The bins are defined at lower
energies starting with the lower energy edge of the first bin set to 0~eV, and
the upper edge to $\sim\ktperp/20$. The edges of subsequent bins are set to
$E_{i+1} = E_i + \, \delta E$. At energies $E_i<\ktperp$ we set the step to
$\delta E \approx E_i$. At higher energies we set the bin width to be
approximately the energy resolution $\Delta E$ given by equation (\ref{eq:res}).
This is motivated by the fact that structures in the cross section which are
separated by less than $\sim\Delta E$ can not be resolved. For energies
$\Ed\gg\Delta E$ the spacing $\Delta \Ed$ between the measured $\alpha_{\rm mb}$
points is often larger than $\Delta E$. For these cases we set the $\sigma_i$
bin edges to fall between the measurement points, i.e, $E_{i} = (E_{{\rm d},i} +
E_{{\rm d},i+1})/2$, so that each  point in $\alpha_{\rm mb}$ effectively
contributes to only one cross section bin.

The binning generated by this method was tested on several DR spectra from
previous TSR measurements for other systems. In all cases a high quality fit was
achieved with reasonable numerical stability. The remaining errors, due to
residual numerical instabilities, basically integrate away when calculating the
plasma rate coefficient $\alpha_{\rm pl}$ using equation (\ref{eq:rateplas}). As
a result, the dominant error in $\alpha_{\rm pl}$ is due to the  statistical and
systematic errors in $\alpha_{\rm mb}$ and the uncertainties in  $\ktperp$ and
$\ktpar$. The resulting uncertainty as it relates to our HCl$^+$ results are
discussed in Section~\ref{sec:res}.


\bibliography{HCl+_DR}


\begin{deluxetable}{cll}
\tablecolumns{3} 
\tablewidth{0pc} 
\tablecaption{ Fit parameters for \HClp DR plasma rate coefficient
$\alpha_{\rm pl}$ using equation~(\ref{eq:plasmafitnew}). 
		\label{tab:plasmares}
		}
\tablehead{ \colhead{Parameter} & \colhead{Value}  & Unit}
\startdata
$A$	& \phs$1.33(4)\e{-8}$	& ${\rm cm^3\ s^{-1}}$ \\
$n$	& \phs$3.7(1)\phn\e{-1}$	& dimensionless\\
$c_1$	& \phs$5.95(4)\e{-4}$	& ${\rm K^{3/2}\ cm^3\ s^{-1}}$ \\
$c_2$	& \phs$1.72(8)\e{-4}$	& ${\rm K^{3/2}\ cm^3\ s^{-1}}$ \\
$c_3$	& $-4.59(4)\e{-4}$	& ${\rm K^{3/2}\ cm^3\ s^{-1}}$ \\
$c_4$	& $-5.3(9)\phn\e{-4}$	& ${\rm K^{3/2}\ cm^3\ s^{-1}}$ \\
$T_1$	& \phs$9.2(2)\phn\e{0}$	& K \\
$T_2$	& \phs$7.7(7)\phn\e{1}$	& K \\
$T_3$	& \phs$8.89(2)\e{1}$	& K \\
$T_4$	& \phs$3.4(4)\phn\e{3}$	& K \\
\enddata
\tablecomments{The values in parentheses give the $1\sigma$ error for the least
significant digit shown.}
\end{deluxetable}

\clearpage


\begin{figure}
\epsscale{0.5}
\plotone{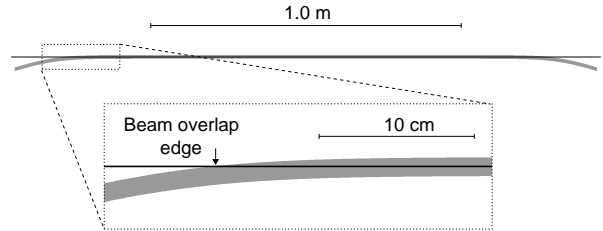}
    \caption{
\doublespace
Top-down view of the geometry in the Target for the electron beam (gray line)
and ion beam (black line). This geometry was used in the Monte-Carlo simulation
to generate the energy distribution $f_{\rm mb}$. The image inset shows the
detail of the electron beam merging region. The system possesses mirror symmetry
around a plane passing through the center of the Target and perpendicular to
both beams. \label{fig:geom}
	}
\end{figure}
\clearpage

\clearpage

\clearpage

\clearpage

\clearpage

\clearpage

\clearpage

\end{document}